\documentclass[journal]{IEEEtran}
\usepackage{amssymb}

\usepackage{subfigure}
\usepackage{graphicx}
\usepackage{enumerate}
\usepackage{amsmath}
\usepackage{amsfonts}
\usepackage{algorithm}
\usepackage{algpseudocode}
\usepackage{url}
\usepackage{epstopdf}
\usepackage{multirow}
\usepackage{authblk}
\usepackage{subfigure}
\usepackage{xcolor}
\usepackage{mathrsfs}

\usepackage{verbatim}

\newtheorem{theorem}{Theorem}[section]
\newtheorem{example}[theorem]{Example}
\newtheorem{corollary}[theorem]{Corollary}
\newtheorem{proposition}[theorem]{Proposition}
\newtheorem{definition}[theorem]{Definition}
\newtheorem{lemma}[theorem]{Lemma}
\newcommand{\prof}{\begin{IEEEproof}}
\newcommand{\eprof}{\end{IEEEproof}}

\newcommand{\prop}{\begin{proposition}}
\newcommand{\eprop}{\end{proposition}}
\newcommand{\them}{\begin{theorem}}
\newcommand{\ethem}{\end{theorem}}
\newcommand{\dfn}{\begin{definition}}
\newcommand{\edfn}{\end{definition}}
\newcommand{\exm}{\begin{example}}
\newcommand{\eexm}{\end{example}}
\newcommand{\coro}{\begin{corollary}}
\newcommand{\ecoro}{\end{corollary}}
\newcommand{\lem}{\begin{lemma}}
\newcommand{\elem}{\end{lemma}}

\newcommand{\eps}{\varepsilon}

\ifCLASSINFOpdf
\else
\fi

\begin{document}

\title{Verification of C-detectability Using Petri Nets}
%
%
%

\author{Hao~Lan, 
        Yin~Tong, ~\IEEEmembership{Member,~IEEE}
        ~Jin~Guo
        and Carla~Seatzu, \IEEEmembership{Senior Member,~IEEE}%
\thanks{H. Lan, Y. Tong (Corresponding Author), and Jin Guo are with the School of Information Science and Technology, Southwest Jiaotong University, Chengdu 611756, China
        {\tt\small haolan@my.swjtu.edu.cn; yintong@swjtu.edu.cn }}
\thanks{C. Seaztu is with the Department of Electrical and Electronic Engineering,
University of Cagliari, 09123 Cagliari, Italy {\tt\small seatzu@diee.unica.it}}}%

%
%

\markboth{Journal of \LaTeX\ Class Files,~Vol.~14, No.~8, August~2015}%
{Shell \MakeLowercase{\textit{et al.}}: Bare Demo of IEEEtran.cls for IEEE Journals}
%



\maketitle
\begin{abstract}
Detectability describes the property of an system whose current and the subsequent states can be uniquely determined after a finite number of observations. In this paper, we relax detectability to \emph{C-detectability} that only requires a given set of crucial states can be distinguished from other states. Four types of C-detectability: strong C-detectability, weak C-detectability, periodically strong C-detectability, and periodically weak C-detectability are defined in the framework of labeled Petri nets, which have larger modeling power than finite automata. Moreover, based on the notion of \emph{basis markings}, the approaches are developed to verify the four C-detectability of a bounded labeled Petri net system. Without computing the whole reachability space and without enumerating all the markings consistent with an observation, the proposed approaches are more efficient.
\end{abstract}


%
\IEEEpeerreviewmaketitle

\section{Introduction}
Due to the limitations of the sensor functions, or the geographical constraints, etc., the system dynamics are usually not perfectly known. However, the state of the system needs to be determined in many applications such as fault diagnosis \cite{cabasino2010fault}, state-feedback control \cite{giua2004observer}, opacity \cite{tong2017verification}, etc. State estimation problems have been widely invested in discrete event systems (DESs) \cite{cabasino2010fault,tong2017verification,lin1988observability,giua2007marking,wu2013comparative,shu2007detectability}.

If the current and the subsequent states of the system can be uniquely determined after the observation of a finite number of events, the system is said to be \emph{detectable}. The notion of detectability was first proposed in \cite{shu2007detectability} in the deterministic finite automaton framework. Four types of detectability have been defined: strong detectability, weak detectability, strong periodic detectability, and weak periodic detectability. It is assumed that the state and the events are partially observable. The approaches to verifying strong/weak detectability and strongly/weak periodic detectability are also presented. The complexity of the approaches are exponential with respect to the number of states of the system. Shu and Lin \cite{shu2011generalized} extend detectability to the nondeterministic finite automata and a more general notion. In \cite{zhang2017problem}, it is shown that the problem of checking weak detectability and weak periodic detectability is PSPACE-complete and that PSPACE-hardness holds even for deterministic DESs with all events observable. Shu and Lin \cite{shu2013delayed} extend strong detectability to delayed DESs and a polynomial algorithm to check strong detectability in delayed DESs is developed. Masopust and Yin \cite{masopust2018complexity} further show that even for very simple DESs that do not have trivial-cycles, the simplest deadlock free DESs, the verification of weak and weak periodic detectability is still intractable.

Petri nets are a graphical and mathematical modeling tool which have higher modeling power than finite automata. Through the structural analysis and applying algebraic techniques, some problems in DESs can be solved more efficiently in Petri nets\cite{tong2017verification,cabasino2011discrete,li2004deadlock}. In \cite{masopust2018deciding}, the authors extend strong detectability and weak detectability in DESs to labeled Petri nets and show that checking strong detectability in labeled Petri nets is EXPSPACE-hard while checking weak detectability is not decidable. For the bounded labeled Petri nets, since their reachability graph (RG) is a finite automaton, the detectability verification problems are clearly decidable. However, it is known that the state explosion issue is unavoidable to construct the RG of large-sized systems. Therefore, applying the automaton based approaches to labeled Petri nets may not work. Meanwhile, detectability is a very restrictive property since it requires that the current and the subsequent states have to be determined without uncertainty while in some cases only a subset of the states are needed to be uniquely identified. 

In the paper, \emph{C-detectability} is defined, where ``C'' stands for ``crucial''. C-detectability requires that if the estimation contains crucial states then the crucial state has to be determined uniquely. Moreover, four types of \emph{C-detectability} are defined in the framework of labeled Petri nets: strong, periodically strong, weak and periodically weak C-detectability. Detectability is a special case of C-detectability where the set of crucial states is equivalent to the state space. Namely, C-detectability is more general than detectability. Based on the notions of \emph{basis markings} and the basis reachability graphs (BRG) efficient approaches to verifying the four C-detectability are proposed. The contributions of the work are summarized as follows.
\begin{itemize}
  \item Strong C-detectability, weak C-detectability, periodically strong C-detectability, and periodically weak C-detectability are formally defined in labeled Petri nets. Compared with detectability defined in \cite{masopust2018deciding}, we assume that the initial state of the observed behavior is not known.
  \item Efficient approaches to verifying the above four C-detectability in bounded labeled Petri nets are proposed. Using basis markings, there is no need to enumerate all the markings that consistent with an observation but solving an integer linear equation. It has been shown that the \emph{basis reachability graph} (BRG) is usually much smaller than the RG. By constructing the observer of the BRG rather than the RG, the four C-detectability can be verified at time.
\end{itemize}

The rest of the paper is organized as follows. In Section~\ref{sec:pre} backgrounds on finite automata, labeled Petri nets and basis markings are recalled. Strong C-detectability, weak C-detectability, periodically strong C-detectability and periodically weak C-detectability in labeled Petri nets are defined in Section~\ref{sec:dect}. In Section~\ref{sec:ver}, the approaches to verifying the four C-detectability in bounded labeled Petri nets is presented. Finally, the paper is concluded and the future work is summarized.

\section{Preliminaries and Background}\label{sec:pre}
In this section we recall the formalism used in the paper and some results on state estimation in Petri nets. For more details, we refer to \cite{cabasino2011discrete,murata1989petri,cassandras2009introduction}.

\subsection{Automata}\label{subsec:auto}
A \emph{nondeterministic finite automaton} (NFA) is a 5-tuple $A=(X, E, f, x_0,X_m)$, where $X$ is the finite \emph{set of states}, $E$ is the finite \emph{set of events}, $f: X\times E\rightarrow 2^X$ is the (partial) \emph{transition relation}, $x_0\in X$ is the \emph{initial state}, and $X_m\subseteq X$ is the set of marked states. If $X_m=\emptyset$, the NFA is denoted as $A=(X, E, f, x_0)$. The transition relation $f$ can be extended to $f:X\times E^*\rightarrow 2^X$ in a standard manner. Given an event sequence $w\in E^*$, if $f(x_0,w)$ is defined in $A$, $f(x_0,w)$ is the set of states reached in $A$ from $x_0$ with $w$ occurring.

Given an NFA, its equivalent DFA, called \emph{observer}, can be constructed following the procedure in Section~2.3.4 of \cite{cassandras2009introduction}. Each state of the observer is a set of states from $X$ that the NFA may be in after an event sequence occurring. Thus, the complexity, in the worst case, of computing the observer is ${\cal O}(2^n)$, where $n$ is the number of states of $A$.

\subsection{Petri Nets}
A \emph{Petri net} is a structure $N=(P,T,Pre,Post)$, where $P$ is a set of $m$ \emph{places}, graphically represented by circles; $T$ is a set of $n$ \emph{transitions}, graphically represented by bars; $Pre:P\times T\rightarrow\mathbb{N}$ and $Post:P\times T\rightarrow\mathbb{N}$ are the \emph{pre-} and \emph{post-incidence functions} that specify the arcs directed from places to transitions, and vice versa\footnote{In this work, we use $\mathbb{N}$, $\mathbb{N}_{\geq 1}$ and $\mathbb{Z}$ to denote the sets of non-negative integers, positive integers and integers, respectively.}. The incidence matrix of a net is denoted by $C=Post-Pre$. A Petri net is said to be \emph{acyclic} if there are no oriented cycles.

A \emph{marking} is a vector $M:P\rightarrow \mathbb{N}$ that assigns to each place a non-negative integer number of tokens, graphically represented by black dots. The marking of place $p$ is denoted by $M(p)$. A marking is also denoted as $M=\sum_{p\in P}M(p)\cdot p$. A \emph{Petri net system} $\langle N,M_0\rangle$ is a net $N$ with \emph{initial marking} $M_0$.

A transition $t$ is \emph{enabled} at marking $M$ if $M\geq Pre(\cdot,t)$ and may fire yielding a new marking $M'=M+C(\cdot,t)$. We write $M[\sigma\rangle$ to denote that the sequence of transitions $\sigma=t_{j1}\cdots t_{jk}$ is enabled at $M$, and $M[\sigma\rangle M'$ to denote that the firing of $\sigma$ yields $M'$. The set of all enabled transition sequences in $N$ from marking $M$ is $L(N,M)=\{\sigma\in T^*| M[\sigma\rangle \}$. Given a sequence $\sigma\in T^*$, the function $\pi:T^*\rightarrow \mathbb{N}^n$ associates with $\sigma$ the Parikh vector $y=\pi(\sigma)\in\mathbb{N}^n$, i.e., $y(t)=k$ if transition $t$ appears $k$ times in $\sigma$. Given a sequence of transitions $\sigma\in T^*$, its \emph{prefix}, denoted as $\sigma'\preceq \sigma$, is a string such that $\exists \sigma''\in T^*:\sigma'\sigma''=\sigma$. The \emph{length} of $\sigma$ is denoted by $|\sigma|$.

A marking $M$ is \emph{reachable} in $\langle N,M_0\rangle$ if there exists a sequence $\sigma$ such that $M_0[\sigma\rangle M$. The set of all markings reachable from $M_0$ defines the \emph{reachability set} of $\langle N,M_0\rangle$, denoted by $R(N,M_0)$. Given a marking $M\in R(N,M_0)$, we define
$$
UR(M)=\{M'\in \mathbb{N}^{m} | M[\sigma_u\rangle M',\sigma_u\in T^*_u\}
$$
its \emph{unobservable reach}, the set of markings reachable from $M$ through unobservable transitions. A Petri net system is \emph{bounded} if there exists a non-negative integer $k \in \mathbb{N}$ such that for any place $p \in P$ and any reachable marking $M \in R(N,M_0)$, $M(p)\leq k$ holds.

A \emph{labeled Petri net} (LPN) is a 4-tuple $G=(N,M_0,\allowbreak E,\ell)$, where $\langle N,M_0\rangle$ is a Petri net system, $E$ is the \emph{alphabet} (a set of labels) and $\ell:T\rightarrow E\cup\{\eps\}$ is the \emph{labeling function} that assigns to each transition $t\in T$ either a symbol from $E$ or the empty word $\eps$. Therefore, the set of transitions can be partitioned into two disjoint sets $T=T_o\cup T_u$, where $T_o=\{t\in T|\ell(t)\in E\}$ is the set of observable transitions and $T_u=T\setminus T_o=\{t\in T|\ell(t)=\eps\}$ is the set of unobservable transitions. The labeling function can be extended to sequences $\ell: T^*\rightarrow E^*$ as $\ell(\sigma t)=\ell(\sigma)\ell(t)$ with $\sigma\in T^*$ and $t\in T$. Given a set $Y\subseteq R(N,M_0)$ of markings, the \emph{language generated by} $G$ \emph{from} $Y$ is ${\cal L}(G,Y)=\bigcup_{M\in Y}\{w\in E^*| \exists \sigma \in L(N,M): w=\ell(\sigma)\}$. In particular, the \emph{language generated by} $G$ is ${\cal L}(G,\{M_0\})=\{w\in E^*|\exists \sigma\in L(N,M_0):w=\ell(\sigma)\}$ that is also simply denoted as ${\cal L}(G)$. Let $w\in {\cal L}(G)$ be an observed word. We define ${\cal C}(w)=\{M\in \mathbb{N}^m|\exists \sigma\in L(N,M_0):M_0[\sigma\rangle M, \ell(\sigma)=w\}$ as the set of markings \emph{consistent} with $w$. When ${\cal C}(w) \neq 1$, markings in ${\cal C}(w)$ are \emph{confusable} with each other as when $w$ is observed it is confused which marking in ${\cal C}(w)$ is the current marking of the system.

Given an LPN $G=(N,M_0,\allowbreak E,\ell)$ and the set of unobservable transitions $T_u$, the \emph{$T_u$-induced subnet} $N'=(P,T',\allowbreak Pre',Post')$ of $N$, is the net resulting by removing all transitions in $T\setminus T_u$ from $N$, where $Pre'$ and $Post'$ are the restriction of $Pre$, $Post$ to $T_u$, respectively. The incidence matrix of the $T_u$-induced subnet is denoted by $C_u=Post'-Pre'$.
\subsection{Basis Markings}\label{sec:basis}
In this subsection we recall some results on state estimation using basis markings proposed in \cite{cabasino2011discrete,tong2017verification}.

\dfn\label{def:exp}
Given a marking $M$ and an observable transition $t\in T_o$, we define $$\Sigma(M,t)=\{\sigma\in T^*_u|M[\sigma\rangle M',M'\geq Pre(\cdot,t)\}$$ as the set of \emph{explanations} of $t$ at $M$ and $Y(M,t)=\{y_u\in \mathbb{N}^{n_u}|\exists \sigma\in \Sigma(M,t):y_u=\pi(\sigma)\}$ the set of \emph{$e$-vectors}. \hfill $\diamond$
\edfn

Thus $\Sigma(M,t)$ is the set of unobservable sequences whose firing at $M$ enables $t$. Among all the explanations, to provide a compact representation of the reachability set we are interested in finding the minimal ones, i.e., the ones whose firing vector is minimal.

\dfn\label{def:minexp}
Given a marking $M$ and an observable transition $t\in T_o$, we define
$$\Sigma_{min}(M,t)=\{\sigma\in \Sigma(M,t)|\nexists \sigma'\in \Sigma(M,t):\pi(\sigma')\lneqq\pi(\sigma)\}$$
as the set of \emph{minimal explanations} of $t$ at $M$ and $Y_{min}(M,t)=\{y_u\in \mathbb{N}^{n_u}|\exists \sigma\in \Sigma_{min}(M,t):y_u=\pi(\sigma)\}$ as the corresponding set of \emph{minimal $e$-vectors}. \hfill $\diamond$
\edfn

Many approaches can be applied to computing $Y_{min}(M,t)$. In particular, when the $T_u$-induced subnet is acyclic the approach proposed by Cabasino \emph{et al}. \cite{cabasino2011discrete} only requires algebraic manipulations. Note that since a given place may have two or more unobservable input transitions, i.e., the $T_u$-induced subnet is not backward conflict free, the set of minimal explanations is not necessarily a singleton.

\dfn\label{def:basisM}
Given an LPN system $G=(N,M_0,E,\ell)$ whose $T_u$-induced subnet is acyclic, its \emph{basis marking set} ${\cal M}_b$ is defined as follows:
\begin{itemize}
  \item $M_0\in {\cal M}_b$;
  \item If $M\in {\cal M}_b$, then $\forall t\in T_o, y_u\in Y_{min}(M,t)$,
  $$M'=M+C(\cdot,t)+C_u\cdot y_u \Rightarrow M'\in {\cal M}_b.$$
\end{itemize}
A marking $M_b\in {\cal M}_b$ is called a \emph{basis marking} of $G$.
\edfn

The set of basis markings contains the initial marking and all other markings that are reachable from a basis marking by firing a sequence $\sigma_u t$, where $t\in T_o$ is an observable transition and $\sigma_u$ is a minimal explanation of $t$ at $M$. Clearly, ${\cal M}_b\subseteq R(N,M_0)$, and in practical cases the number of
basis markings is much smaller than the number of reachable markings \cite{cabasino2011discrete,tong2017verification,ziyue2017basis}. We denote $${\cal C}_b(w)={\cal M}_b\cap {\cal C}(w)$$ the set of basis markings consistent with a given observation $w\in {\cal L}(G)$.

\begin{figure}
  \centering
  \includegraphics[width=0.35\textwidth]{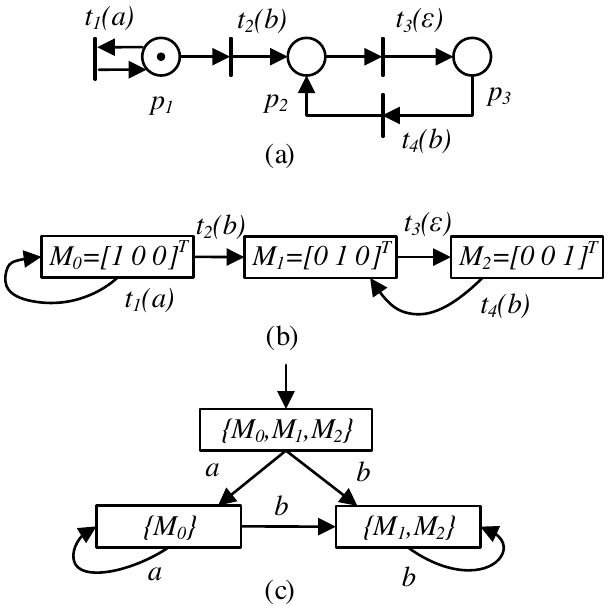}\\
  \caption{An LPN system (a), its RG (b), and the Observer of the RG (c).}\label{fig:case1}
\end{figure}

\exm\label{eg:miniExp}
Consider the LPN system in Fig.~\ref{fig:case1}(a), where transition $t_3$ is unobservable, and transition $t_1$ is labeled by $a$, $t_2$ and $t_4$ are labeled by $b$. At the initial marking $M_0=[1\ 0\ 0]^T$, the minimal explanations of $t_2$ is $\Sigma_{min}(M_0,t_2)=\{\eps\}$, and thus $Y_{min}(M_0,t_2)=\{\vec{0}\}$. The corresponding basis marking is $M_0+C(\cdot,t_2)+C_u\cdot \vec{0}=M_1=[0\ 1\ 0]^T$. At $M_1$, the minimal explanation of $t_4$ is $\Sigma_{min}(M_1,t_4)=\{t_3\}$, and thus $Y_{min}(M_1,t_4)=\{[1]^T\}$. The basis marking obtained is $M_1+C(\cdot,t_4)+C_u\cdot [1]^T=M_1$.  \hfill $\diamond$
\eexm

\prop\label{prop:basis}\cite{cabasino2011discrete}
Let $G=(N,M_0,E,\ell)$ be an LPN system whose $T_u$-induced subnet is acyclic, $M_b\in {\cal M}_b$ a basis marking of $G$, and $w\in {\cal L}(G)$ an observation generated by $G$. We have
\begin{enumerate}
  \item a marking $M$ is reachable from $M_b$ iff
\begin{equation}\label{eq:basis}
M=M_b+C_u\cdot y_u
\end{equation}
has a nonnegative solution $y_u\in \mathbb{N}^{n_u}$.
  \item \begin{align}\label{eq:cw}
  {\cal C}(w)= \bigcup_{ M_b\in {\cal C}_b(w)} & UR(M_b)  \nonumber\\
                \bigcup_{ M_b\in {\cal C}_b(w)} & \{M\in \mathbb{N}^m|M=M_b+C_u\cdot y_u,\nonumber\\
  & y_u\in  \mathbb{N}^{n_u}\}.
    \end{align}
\end{enumerate}
\eprop

Statement 1) of Proposition~\ref{prop:basis} implies that any solution $y_u \in \mathbb{N}^{n_u}$ of Eq.~\eqref{eq:basis} corresponds to the firing vector of a firable sequence $\sigma$ from $M_b$, i.e., $M_b[\sigma\rangle$ and $\pi(\sigma)=y_u$. According to Statement 2), the set of markings consistent with an observation can be characterized using linear algebra without an exhaustive marking enumeration.


\section{C-Detectability}\label{sec:dect}
Detectability is a property that the current and the subsequent states of the system can be uniquely determined after a finite delayed of observations. However, the requirement of exactly determining the state is too strong in some applications. For instance, opacity only determines whether secret states confuse with non-secret states. In the paper, we relax detectability to \emph{C-detectability}, where ``C'' stands for ``crucial'' and that requires only if a given set of states, called crucial states, can be distinguished from other states.

The following two assumptions are made: 1) the LPN system $G$ is deadlock free. Namely, for any marking of the LPN there exists at least one transition enabled: $\forall M\in R(N,M_0), \exists t\in T$ such that $M[t\rangle$; 2) the $T_u$-induced subnet is acyclic.

Assumption 1) is made to avoid unnecessarily complicated technicalities developed in the paper. Assumption 2) makes sure that eventually an observable transition will occur and produce an observation. These two assumptions guarantee that any transition sequence fired by the system can continue infinitely and the corresponding observation can also be infinite long. Note that Assumption 2) is stronger than assuming there are no infinite
strings of unobservable transitions in the automata framework since the existence of cycles of the unobservable transitions in the net structure does not imply that such a cycle is enabled. Assumption 2) is a structural assumption that can be verified in polynomial time and it brings the computational advantages for the verification of C-detectability (as showing in the next section).

Note that in the framework of automata, it is assumed that the initial state of the system is not known. However, in the Petri net framework, a net structure $N$ without an initial marking $M_0$ is not even a dynamic system. The LPN counterpart is to assume that the net system $(N,M_0,E,\ell)$ is known but it is not known from which marking the observation is generated. Namely, an observation could be generated from any marking in $R(N,M_0)$.

Therefore, the language generated by $G$ is extended to ${\cal L}(G)={\cal L}(G,R(N,M_0))$.  Let $w\in {\cal L}(G)$ be an observed word. We extend the set of marking consistent with $w$ to $$
\begin{aligned}
{\cal C}(w)=\{M\in \mathbb{N}^m|&\exists M'\in R(N,M_0), \sigma\in L(N,M'),:\\
& M'[\sigma\rangle M, \ell(\sigma)=w\}.
\end{aligned}$$
Note that the extensions of ${\cal L}(G)$ and ${\cal C}(w)$ will not change the results in Proposition~\ref{prop:basis}. Correspondingly, we define $L(G)=\{\sigma\in T^*| \exists M\in R(N,M_0): M[\sigma\rangle\}$ the set of transition sequences enabled at a marking in $R(N,M_0)$.  We denote $$L^{\omega}(G)=\{\sigma\in T^*|\sigma\in L(G) \wedge |\sigma|\text{ is infinite}\}$$ the set of transition sequences of infinite length that are enabled by a marking of $G$.

\dfn\label{def:Mc}
Given an LPN system $G$, the set of \emph{crucial markings} is defined as a subset of markings ${\cal M}_c \subseteq R(N,M_0)$ of $G$.
\edfn

C-detectability requires that any crucial marking $M\in{\cal M}_c$ is distinguishable from other markings after a finite number of observations. Four C-detectability in LPNs are formally defined.

\dfn\label{def:S-C-detect}
[Strong C-detectability] Let $G=(N,M_0,\allowbreak E,\ell)$ be an LPN system and ${\cal M}_c$ the set of crucial markings. System $G$ is said to be \emph{strongly C-detectable} with respect to ${\cal M}_c$ if there exists a finite integer $K\in \mathbb{N}$ such that $\forall \sigma\in L^\omega(G)$, $\forall \sigma'\preceq \sigma$, $|w|\geq K$,
\begin{equation}\label{eq:C-detect}
{\cal C}(w)\cap {\cal M}_c \neq \emptyset \Rightarrow |{\cal C}(w)|=1,
\end{equation}
where $w=\ell(\sigma')$.\hfill $\diamond$
\edfn

An LPN system is strongly C-detectable if we can uniquely determine the markings in ${\cal M}_c$ \emph{all} the time after a finite number of observations for \emph{all} trajectories of the system. On the contrary, if there is no crucial marking in ${\cal C}(w)$ it does not mater whether marking in ${\cal C}(w)$ can be distinguished from other markings or not.

\exm\label{eg:SCD}
Consider the LPN system in Fig.~\ref{fig:case1}(a). Let the set of crucial markings be ${\cal M}_c =\{[1\ 0\ 0]^T\}$.
Its RG is shown in Fig.~\ref{fig:case1}(b). Since the initial marking from which the observation started is assumed unknown, the initial state of the observer is $R(N,M_0)$ rather than $M_0$ (see Fig.~\ref{fig:case1}(c)). After $a^*$ is observed, the current state of the system can be uniquely determined, namely $M_0$. If $b^*$ is observed, the estimation of the current marking is ${\cal C}(b^*)=\{M_1,M_2\}$. The system is considered to be C-detectable since there is no crucial marking in the estimation. Therefore, by Definition~\ref{def:S-C-detect}, the LPN system is strongly C-detectable w.r.t ${\cal M}_c$. \hfill $\diamond$
\eexm

\dfn\label{def:C-detect}
[Weak C-detectability] Let $G=(N,M_0,\allowbreak E,\ell)$ be an LPN system and ${\cal M}_c$ the set of  crucial markings. System $G$ is said to be \emph{weakly C-detectable} with respect to ${\cal M}_c$ if there exists a finite integer $K\in \mathbb{N}$ such that $\exists \sigma\in  L^\omega(G)$, $\forall \sigma'\preceq \sigma$, $|w|\geq K$,
$${\cal C}(w)\cap {\cal M}_c \neq \emptyset \Rightarrow |{\cal C}(w)|=1,$$
where $w=\ell(\sigma')$. \hfill $\diamond$
\edfn

Compared with strong C-detectability, weak C-detectability only requires that the markings in ${\cal M}_c$ can be distinguished from others \emph{all} the time, after a finite number of observations, for \emph{some} trajectories, rather than all the trajectories of the system. According to the definitions of strong C-detectability and weak C-detectability, obviously we have if an LPN system is strongly C-detectable, it is weakly C-detectable as well.

\exm\label{eg:WD}
Let us consider again the LPN system in Fig.~\ref{fig:case1}(a). Let the set of crucial markings be ${\cal M}_c =\{[1\ 0\ 0]^T,[0\ 1\ 0]^T\}$. Clearly, the LPN system is not strongly C-detectable w.r.t ${\cal M}_c$ since after $b^*$ is observed, the estimation contains a crucial marking $M_1$ which cannot be distinguished from $M_2$. Therefore, by Definitions~\ref{def:C-detect}, the LPN system is only weakly C-detectable w.r.t ${\cal M}_c$. \hfill $\diamond$
\eexm

\dfn\label{def:S-P-C-detect}
[Periodically strong C-detectability] Let $G=(N,M_0,\allowbreak E,\ell)$ be an LPN system and ${\cal M}_c$ the set of crucial markings. System $G$ is said to be \emph{periodically strongly C-detectable} with respect to ${\cal M}_c$ if there exists a finite integer $K\in \mathbb{N}$ such that $\forall \sigma\in L^\omega (G),\forall \sigma' \preceq \sigma$,
\begin{align}\label{eq:pd}
\exists \sigma''\in T^*: & \sigma'\sigma''\preceq \sigma, |\ell(\sigma'')|<K,\text{ and}  \nonumber \\
&{\cal C}(w)\cap {\cal M}_c \neq \emptyset  \Rightarrow |{\cal C}(w)|=1,
\end{align}
where $w=\ell(\sigma'\sigma'')$. \hfill $\diamond$
\edfn

An LPN system is periodically strongly C-detectable if we can periodically distinguish markings in ${\cal M}_c$ for \emph{all} trajectories of the system. In other words, ``periodically C-detectable'' means as a string $\sigma$ continues \emph{eventually} there is only one marking (or no marking) in ${\cal M}_c$ consistent with the corresponding observation. We point out that for different evolutions the period may be different. Note that as the system is bounded one can find an upper bound for all periods.

\begin{figure}
  \centering
  \includegraphics[width=0.45\textwidth]{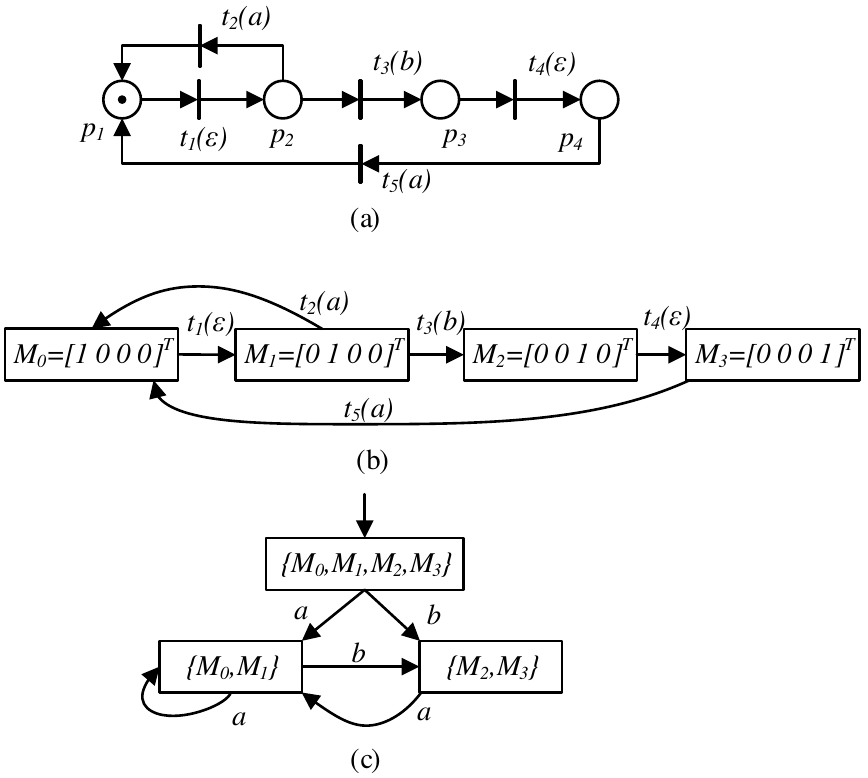}\\
  \caption{LPN system in Example~\ref{eg:PSD} (a), its RG (b), and the observer of the RG (c).}\label{fig:case2}
\end{figure}

\exm\label{eg:PSD}
Consider the LPN system in Fig.~\ref{fig:case2}(a). Let the set of crucial markings be ${\cal M}_c =\{[0\ 0\ 1\ 0]^T\}$.
Its RG is shown in Fig.~\ref{fig:case2}(b), and the observer of the RG is shown in Fig.~\ref{fig:case2}(c). In state $\{M_0,M_1\}$ of the observer, there is no crucial marking in the estimation, i.e., no crucial marking is confused with other markings. Therefore, for observation $(ba)^*$, the crucial marking is considered periodically distinguished since there exists a state $\{M_2,M_3\}$ such that the crucial marking $M_2$ cannot be distinguished from $M_3$. By Definition~\ref{def:S-P-C-detect}, the LPN system is periodically strongly C-detectable. \hfill $\diamond$
\eexm

\dfn\label{def:P-C-detect}
[Periodically weak C-detectability] Let $G=(N,M_0,\allowbreak E,\ell)$ be an LPN system and ${\cal M}_c$ the set of crucial markings. System $G$ is said to be \emph{periodically weakly C-detectable} with respect to ${\cal M}_c$ if there exists a finite integer $K\in \mathbb{N}$ such that $\exists \sigma\in L^\omega (G),\forall \sigma' \preceq \sigma$,
$$\begin{aligned}
\exists \sigma''\in T^*: & \sigma'\sigma''\preceq \sigma, |\ell(\sigma'')|<K,\\
&{\cal C}(w)\cap {\cal M}_c \neq \emptyset \Rightarrow |{\cal C}(w)|=1,
\end{aligned}$$
where $w=\ell(\sigma'\sigma'')$. \hfill $\diamond$
\edfn

An LPN system is periodically weakly C-detectable if we can periodically distinguish markings in ${\cal M}_c$ for \emph{some} trajectories of the system.

\exm\label{eg:PWD}
Let us consider again the LPN system in Fig.~\ref{fig:case2}(a). Let the set of crucial markings be ${\cal M}_c =\{[1\ 0\ 0\ 0]^T\}$.
When $a^*$ is observed, the crucial marking $M_0$ is confused with $M_1$ while if $(ba)^*$ is observed, the crucial marking can be distinguished periodically since state $\{M_2,M_3\}$ does not contain any crucial marking. By Definition~\ref{def:P-C-detect}, the LPN system is periodically weakly C-detectable. \hfill $\diamond$
\eexm

By Definitions \ref{def:S-C-detect} to \ref{def:P-C-detect}, if an LPN system is periodically
strongly C-detectable, it is also periodically weakly C-detectable;
if an LPN system is strongly (resp., weakly) C-detectable, it is periodically strongly (resp., weakly) C-detectable as well.

\section{Verifying C-detectability}\label{sec:ver}

In the automaton framework, an observer is constructed to verify detectability \cite{shu2007detectability,shu2011generalized}. To verify detectability of bounded LPNs, first one needs to construct the RG and then to follow the methods by constructing the observer of the RG (as Examples~\ref{eg:SCD} to \ref{eg:PWD} showing). It is known that, the complexity of constructing the RG of a Petri net system is exponential to its size\footnote{The size of a Petri net system usually refers the number of places, transitions, and the initial tokens, etc.}. Moreover, in the worst case, the complexity of constructing the observer is exponential to the number of states of the system. Therefore, to verify detectability of large-scaled systems the state explosion problem cannot be avoided. In this section, we present an approach to verifying four C-detectability without enumerating all states of the system. In this way, the state explosion problem is practically avoided.

\subsection{Construction of the BRG}
Based on the notion of basis markings, we introduce the \emph{basis reachability graph} (BRG) for C-detectability. To guarantee that the BRG is finite, we assume that the LPN system is bounded.


For each basis marking $M_b\in {\cal M}_b$ two values are respectively assigned by functions $\alpha:{\cal M}_b\rightarrow \{0,1\}$ and $\beta:{\cal M}_b\rightarrow \{0,1\}$ that are defined by Eqs.~\eqref{eq:alpha} and \eqref{eq:beta}.

\begin{equation}\label{eq:alpha}
\alpha(M_b)=\left\{\begin{array}{ll}
  1 & \text{if ${\cal M}_c\cap UR(M_{b})\neq \emptyset$;}\\
  0 & \text{otherwise.}
\end{array}\right.
\end{equation}
\begin{equation}\label{eq:beta}
\beta(M_b)=\left\{\begin{array}{ll}
  1 & \text{$|UR(M_{b})|>1$;}\\
  0 & \text{$|UR(M_{b})|=1$.}
\end{array}\right.
\end{equation}
Provided that the $T_u$-induced subnet is acyclic, $UR(M_b)$ can be re-written as
\begin{equation}\label{eq:ur}
  UR(M_b)=\{M\in \mathbb{N}^m|M=M_b+C_u\cdot y_u, y_u\in \mathbb{N}^{n_u}\}.
\end{equation}
Since $M_b\in UR(M_b)$, i.e., $y_u=\vec{0}$ is a solution to Eq.~\eqref{eq:ur}, $|UR(M_b)|\geq 1$. In addition, $|UR(M_b)|>1$ iff Eq.~\eqref{eq:ur} has a positive integer solution $y_u\in \mathbb{N}^{n_u}_{\geq 1}$.

We denote $B = (X, E, f, x_0)$ the BRG for C-detectability of an LPN system $G = (N, M_0, E, \ell)$, where $X\subseteq {\cal M}_b\times \{0,1\}\times \{0,1\}$ is a finite set of states, and each state $x\in X$ of the BRG is a triple $(M_b,\alpha(M_b), \beta(M_b))$. We denote the $i$-th (with $i=1,2,3$) element of $x$ as $x(i)$. The initial node of the BRG is $x_0=(M_0,\alpha(M_0), \beta(M_0))$. The event set of the BRG is identical to the alphabet $E$. The transition function $f: X\times E \rightarrow X$ can be determined by the following rule. If at marking $M_{b}$ there is an observable transition $t$ for which a minimal explanation exists and the firing of $t$ and one of its minimal explanations lead to $M_{b}'$, then an edge from node $(M_{b},\alpha(M_{b}), \beta(M_{b}))$ to node $(M_{b}',\alpha(M_{b}'), \beta(M_{b}'))$ labeled with $\ell(t)$ is defined in the BRG. The procedure of constructing the BRG for C-detectability is summarized in Algorithm~\ref{alo:BRG}.

\begin{algorithm}
\caption{Computation of the BRG for C-detectability}
\label{alo:BRG}
  \begin{algorithmic}[1]
   \Require
    A bounded LPN system $G=(N,M_0,E,\ell)$, and the set of crucial markings ${\cal M}_c$.
   \Ensure
    The corresponding BRG $B = (X, E, f, x_0)$.
   \State $x_0:=(M_0,\alpha(M_0),\beta(M_0))$.
   \State $X:=\{x_0\}, X_{new}:=\{x_0\}$.
   \ForAll{nodes $x\in X_{new}$,}
    \State $M:=x(1)$;
    \ForAll{$t$ s.t. $Y_{min}(M,t)\neq \emptyset$}
     \ForAll {$e\in Y_{min}(M,t)$}
      \State $M':=M+C_u\cdot e+C(\cdot,t)$;
      \State compute $\alpha(M')$ and $\beta(M')$;
      \State $x'=(M',\alpha(M'),\beta(M'))$;
      \State $f(x,\ell(t))=x'$;
      \If {$x'\notin X$,}
       \State $X=X\cup \{x'\}$, $X_{new}=X_{new}\cup \{x'\}$;
      \EndIf
     \EndFor
    \EndFor
    \State $X_{new}=X_{new} \setminus \{x\}$.
   \EndFor
  \end{algorithmic}
\end{algorithm}

\lem\label{lem:alpha}
Let $G$ be an LPN system, ${\cal M}_c$ the set of crucial markings, and $M_b\in {\cal M}_{b}$ a basis marking. If $\alpha(M_b)=1$, there exists an observation $w$ such that ${\cal C}(w)\cap {\cal M}_c\neq \emptyset$.
\elem
\prof
By assumption $\alpha(M_b)=1$ in Eq.~\ref{eq:alpha}, there is a marking $M\in {\cal M}_c\cap UR(M_{b})$. Let $\sigma\in T^*$ be the transition sequence such that $M_0[\sigma\rangle M_b$ and $\ell(\sigma)=w$. Clearly, $M_0[\sigma\rangle M_b[\sigma_u\rangle M$, where $\sigma_u\in T_u^*$, and $\ell(\sigma\sigma_u)=w$. Therefore, $M\in {\cal C}(w)$ and $M\in {\cal C}(w)\cap {\cal M}_c\neq \emptyset$.
\eprof

In simple words, if $\alpha(M_b)=1$, there exists an observation $w$ such that ${\cal C}(w)$ contain crucial markings. However, if $\alpha(M_b)=0$ there may exist an observation $w$ and another basis marking $M'_b\neq M_b$ such that $M_b', M_b\in {\cal C}(w)$ but $\alpha(M_b')=1$. In this case, ${\cal C}(w)\cap {\cal M}_c$ is still not empty.

\lem\label{lem:beta}
Let $G$ be an LPN system, ${\cal M}_c$ the set of crucial markings, and $M_b\in {\cal M}_{b}$ a basis marking. If $\beta(M_b)=1$, there exists an observation $w$ such that $|{\cal C}(w)|> 1$.
\elem

\prof
Let $\sigma\in T^*$ be the transition sequence that $M_0[\sigma\rangle M_b$ and $\ell(\sigma)=w$. Since $UR(M_b)\subseteq {\cal C}(w)$ and by assumption $\beta(M_b)=1$, i.e., $|UR(M_b)|>1$, $|{\cal C}(w)|\neq 1$.
\eprof

In simple words, if $\beta(M_b)=1$, there is an observation $w$ such that ${\cal C}(w)$ contains more than one marking. However, if $\beta(M_b)=0$ there may be another basis marking $M_b'$ such that $M_b,M_b'\in {\cal C}(w)$ and $M_b'\notin UR(M_b)$. In this case, $|{\cal C}(w)|$ is still greater than 1.

%

Lemmas~\ref{lem:alpha} and \ref{lem:beta} show that constructing the BRG is not sufficient for C-detectability analysis. Next we construct the observer of the BRG to further investigate the sufficient and necessary conditions for C-detectability.

We denote $B_o=({\cal{X}},E,\delta,\hat{X}_0,{\cal X}_m)$ the observer of the BRG $B = (X, E, f, x_0)$ for C-detectability, where ${\cal{X}}\subseteq 2^X$ is a finite set of nodes and ${\cal X}_m\subseteq {\cal X}$ is the set of marked states. The event set of the observer is identical to the alphabet $E$. The transition function is $\delta: {\cal{X}}\times E \rightarrow {\cal{X}}$. Since it is assumed that the marking from which the observation generated is not known, the initial state $\hat{X}_0={\cal M}_{b}$. Clearly, all markings in a state of $B_o$ correspond to markings in a set ${\cal C}_b(w)$. Namely, let $\delta({\cal M}_{b},w)=\hat{X}$, then ${\cal C}_b(w)=\bigcup_{x\in \hat{X}} x(1)$. The observer of the BRG can be constructed by applying the standard determization algorithm in \cite{cassandras2008introduction}.

\prop\label{prop:alpha}
Let $G$ be an LPN system, $B_o=({\cal{X}},E,\delta,\hat{X}_0,{\cal X}_m)$ the observer of its BRG, and ${\cal M}_c$ the set of crucial markings. For $\hat{X}\in {\cal X}$, iff $\forall x\in \hat{X}, x(2)=0$, there exists an observation $w$ such that ${\cal C}(w)\cap {\cal M}_c= \emptyset$.
\eprop
\prof
Let $w\in E^*$ be an observation and $\delta({\cal M}_{b},w)=\hat{X}$. Since by assumption for all $x\in \hat X$, $x(2)=0$, i.e., for all $M_b\in {\cal C}_b(w)$, $UR(M_b)\cap {\cal M}_c=\emptyset$, and by Eq.~\eqref{eq:cw}, ${\cal C}(w)=\bigcup_{M_b\in {\cal C}_b(w)} UR(M_b)$, ${\cal C}(w)\cap {\cal M}_c= \emptyset$.
\eprof

In simple words, given a state $\hat{X}\in \mathcal{X}$, if $\hat{X}$ have all the triple $(M_{b},\alpha(M_{b}), \beta(M_{b}))$ with $\alpha(M_{b})=0$, there exists an observation $w$ such that ${\cal C}(w)$ does not contain any crucial markings.

\coro
Let $G$ be an LPN system and $B_o=({\cal{X}},E,\delta,\hat{X}_0,{\cal X}_m)$ the observer of its BRG. If $\forall \hat X\in {\cal X}, \forall x\in \hat X$, $x(2)=0$, the LPN system is strongly C-detectable, weakly C-detectable, periodically strongly C-detectable, and periodically weakly C-detectable.
\ecoro

\prop\label{prop:beta}
Let $G$ be an LPN system and $B_o=({\cal{X}},E,\delta,\hat{X}_0,{\cal X}_m)$ the observer of its BRG. Given a state $\hat{X}\in {\cal X} $, iff $\hat{X}=\{(M_b,\alpha(M_{b}),0)\}$, where $M_b$ is a basis marking of $G$, there exists an observation $w$ such that $|{\cal C}(w)|=1$.
\eprop
\prof
By assumption $\hat{X}=\{(M_b,\alpha(M_{b}),0)\}$, there exists an observation $w$ such that $\delta({\cal M}_{b},w)=\hat{X}$, and ${\cal C}_b(w)=\{M_b\}$. Since $\beta(M_b)=0$, i.e., $UR(M_b)=\{M_b\}$, and by Eq.~\eqref{eq:cw}, ${\cal C}(w)={\cal C}_b(w)=\{M_b\}$ and $|{\cal C}(w)|=1$.
\eprof

In simple words, given a state $\hat{X}\in \mathcal{X}$, if $\hat{X}$ contains only one state $x=(M_{b},\alpha(M_{b}), \beta(M_{b}))$ with $\beta(M_b)=0$, there is an observation $w$ whose corresponding ${\cal C}(w)$ contains only one marking $M_b$.

\coro
Let $G$ be an LPN system and $B_o=({\cal{X}},E,\delta,\hat{X}_0,{\cal X}_m)$ the observer of its BRG. If all states $\hat X\in {\cal X}$ of $B_o$ has the form $\hat{X}=\{(M_b,\alpha(M_{b}),0)\}$, the LPN system is strongly C-detectable, weakly C-detectable, periodically strongly C-detectable, and periodically weakly C-detectable.
\ecoro

Next we define the set of marked states as the set of states in $B_o$ that contain only one or no crucial marking:
\begin{align*}
\mathcal{X}_m=& \{\hat{X}\in \mathcal{X}|\hat{X}=\{(M_b,\alpha(M_{b}),0)\}\} \cup\\
      & \{\hat{X}\in \mathcal{X}|\forall x\in \hat{X}, x(2)=0\}.
\end{align*}

\prop\label{prop:Xm}
Let $G$ be an LPN system, $B_o=({\cal{X}},E,\delta,\hat{X}_0,{\cal X}_m)$ the observer of its BRG, and ${\cal M}_c$ the set of crucial markings. Given a state $\hat{X}\in {\cal X}$, iff $\hat{X}\in {\cal X}_m$, there exists an observation $w$ such that ${\cal C}(w)\cap {\cal M}_c \neq \emptyset \Rightarrow |{\cal C}(w)|=1$.
\eprop
\prof
Follows from Propositions~\ref{prop:alpha} and \ref{prop:beta}.
\eprof
Namely, given an observation that leads to a marked state in the observer its current marking can be distinguished from a crucial marking.

\begin{figure}
  \centering
  \includegraphics[width=0.5\textwidth]{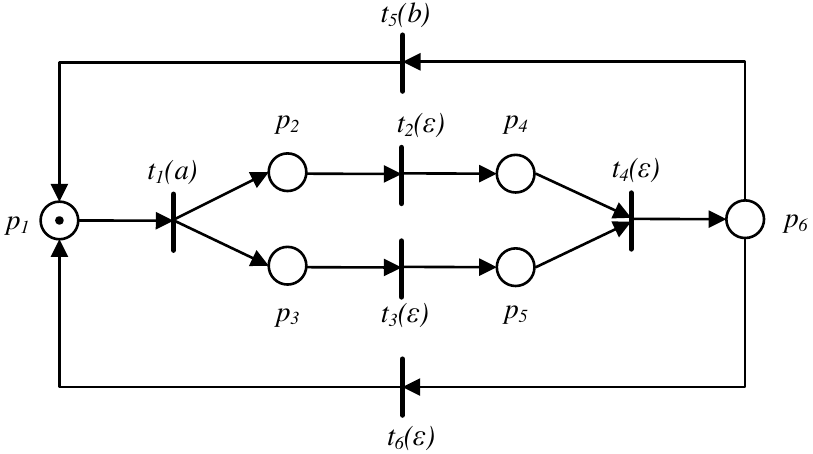}\\
  \caption{The LPN system in Example~\ref{eg:obs-BRG}.}\label{fig:LPN}
\end{figure}

\begin{figure}
  \centering
  \includegraphics[width=0.2\textwidth]{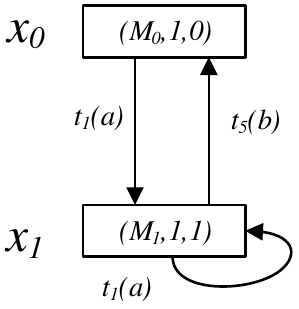}\\
  \caption{BRG of the LPN in Fig.~\ref{fig:LPN}}\label{fig:BRG}
\end{figure}

\begin{figure}
  \centering
  \includegraphics[width=0.35\textwidth]{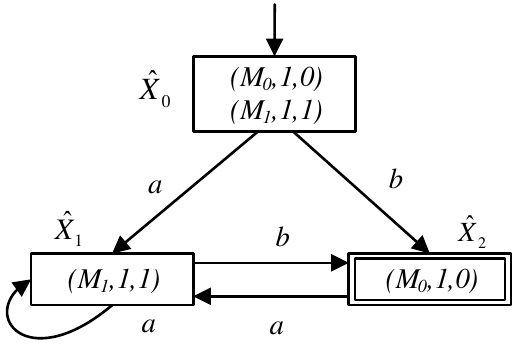}\\
  \caption{Observer of the BRG in Fig.~\ref{fig:BRG}}\label{fig:observer}
\end{figure}

\exm\label{eg:obs-BRG}
Let us consider the LPN system in Fig.~\ref{fig:LPN} whose $T_u$-induced subnet is acyclic and where $T_o=\{t_1, t_5\}$, $T_u=\{t_2, t_3, t_4, t_6\}$. Let the set of crucial markings be ${\cal M}_c =\{[1\ 0\ 0\ 0\ 0\ 0]^T\}$. The LPN system has 6 reachable markings while 2 basis markings $M_0=[1\ 0\ 0\ 0\ 0\ 0]^T$ and $M_1=[0\ 1\ 1\ 0\ 0\ 0]^T$. For $M_0$, by Eq.~\eqref{eq:ur} $UR(M_0)=\{M_0\}$, ${\cal M}_c\cap UR(M_0)\neq \emptyset$ and $|UR(M_0)|=1$. Therefore, $\alpha(M_0)=1$ and $\beta(M_0)=0$ by Eq.~\eqref{eq:alpha} and Eq.~\eqref{eq:beta}. For $M_1$, $UR(M_1)=R(N,M_0)$, ${\cal M}_c\cap UR(M_1)\neq \emptyset$ and $|UR(M_1)|>1$. Therefore, $\alpha(M_1)=1$ and $\beta(M_1)=1$.
By Algorithm~\ref{alo:BRG}, its BRG for C-detectability is presented in Fig.~\ref{fig:BRG}. Since the initial state is assumed to be unknown, it is not specified in the BRG. The observer of the BRG for C-detectability is shown in Fig.~\ref{fig:observer}. By definition, $\mathcal{X}_m=\{\hat{X}_2\}$. \hfill $\diamond$
\eexm

\subsection{Verification of C-detectability}
Since C-detectability considers the transition sequences of infinite length, we first study the properties of cycles in the observer.

\dfn\label{def:cycle}[Simple cycle]
A \emph{(simple) cycle} in the observer $B_o=({\cal{X}},E,\delta,\hat{X}_0,{\cal X}_m)$ of a BRG is a path $\gamma_j=\hat X_{j1}e_{j1}\hat X_{j2}\ldots \hat X_{jk}\allowbreak e_{jk}\hat X_{j1}$ that starts and ends at the same state but without repeat edges, where $\hat X_{ji}\in {\cal X}$ and $e_{ji}\in E$ with $i=\{1,2,\ldots,k\}$, and $\forall m,n\in\{1,2,3,\ldots,k\}$ with $m\neq n$, $\hat X_{jm}\neq \hat X_{jn}$. The corresponding observation of the cycle is $w=e_{j1}\ldots e_{jk}$. A state $\hat X_{ji}$ contained in $\gamma_j$ is denoted by $\hat X_{ji}\in \gamma_j$. The set of simple cycles in the observer is denoted by $\Gamma$. \hfill $\diamond$
\edfn

\dfn\label{def:unamb}
[Unambiguous cycle] A cycle $\gamma_j\in \Gamma$ of the observer $B_o=({\cal{X}},E,\delta,\hat{X}_0,{\cal X}_m)$ is said to be
\begin{itemize}
  \item \emph{unambiguous} w.r.t ${\cal M}_c$ if $\forall \hat X\in \gamma_j$, $\hat X\in {\cal X}_m$.
  \item \emph{semi-unambiguous} w.r.t ${\cal M}_c$ if $\exists \hat X\in \gamma_j$, $\hat X\in {\cal X}_m$.
  \item \emph{ambiguous} w.r.t ${\cal M}_c$ if $\forall \hat X\in \gamma_j$, $\hat X\notin {\cal X}_m$. \hfill $\diamond$
\end{itemize}
\edfn

If all states contained in a cycle are marked, then the cycle is called an unambiguous cycle since the current marking of the corresponding observation of the cycle can be distinguished from the crucial markings. On the contrary, if there exists at least one marked state in the cycle, then it is called semi-unambiguous. Otherwise, the cycle is said to be ambiguous because the current marking is always confused with others. We prove that for an unambiguous cycle, any transition sequence consistent with the observation of the cycle, it also satisfies the condition in Eq.~\eqref{eq:C-detect}.

\prop\label{prop:unamb}
Let $G=(N,M_0,\allowbreak E,\ell)$ be an LPN system, and ${\cal M}_c$ the set of crucial markings. There exists an unambiguous cycle $\gamma_j$ in the observer of its BRG iff there exists a finite $K\in \mathbb{N}$ such that $\exists \sigma\in L^\omega(G)$ with $\ell(\sigma)=w$, where $w$ is the corresponding observation of $\gamma_j$, $\forall \sigma_1\preceq \sigma$, $\ell(\sigma_1)=w_1$, $|w_1|\geq K$, the condition ${\cal C}(w_1)\cap {\cal M}_c \neq \emptyset \Rightarrow |{\cal C}(w_1)|=1$ holds.
\eprop
\prof
(If) Assume $\exists K\in \mathbb{N}$, $\exists \sigma\in L^\omega(G)$ such that $\forall \sigma_1\preceq \sigma$ with $\ell(\sigma_1)=w_1\preceq w$, $|w_1|\geq K$, ${\cal C}(w_1)\cap {\cal M}_c \neq \emptyset \Rightarrow |{\cal C}(w_1)|=1$.
Since $\sigma$ is of an infinite length and $B_o$ has a finite number of nodes, the path along $\ell(\sigma)=w$ must contain a cycle $\gamma_j=\hat X_{j1}e_{j1}\hat X_{j2}\ldots \hat X_{jk}\allowbreak e_{jk}\hat X_{j1}\in \Gamma$, i.e., there exist $w_0\in E^*$ and $w_2\in \{e_{j1},\ldots,e_{jk}\}^*$ such that $w=w_0(e_{j1}\ldots e_{jk})^*w_2$ and $|w_0|$ is finite. Let $|w_0|=K$.
Since ${\cal C}(w_1)\cap {\cal M}_c \neq \emptyset \Rightarrow |{\cal C}(w_1)|=1$, by Proposition~\ref{prop:Xm}, $\delta({\cal M}_b,w_1)\in {\cal X}_m$. By assumption for all $w_1\preceq w$ with $|w_1|\geq K$ (i.e., $|w_1|\geq |w_0|$) the condition ${\cal C}(w_1)\cap {\cal M}_c \neq \emptyset \Rightarrow |{\cal C}(w_1)|=1$ holds, $\forall \hat X\in \gamma_j$, $\hat X \in \delta({\cal M}_b,w_1) \in{\cal X}_m$. Thus the cycle $\gamma_j$ is an unambiguous cycle.

(Only if) Assume that exists an unambiguous cycle $\gamma_j=\hat X_{j1}e_{j1}\hat X_{j2}\ldots \hat X_{jk}\allowbreak e_{jk}\hat X_{j1}\in \Gamma: \forall \hat X\in \gamma_j$, $\hat X\in {\cal X}_m$.
 Clearly, there exists $\sigma\in L^\omega(G)$ and $w_0\in E^*$ such that $\ell(\sigma)=w=w_0(e_{j1}\ldots e_{jk})^*$ with $|w_0|$ is finite. Let $|w_0| = K \in\mathbb{N}$.
 Since the $T_u$-induced subnet is acyclic, for all $\sigma_1\in L(G)$ with $\ell(\sigma_1)=w_1$, $|w_1|\geq |w_0|$ (i.e.,$|w_1|\geq K$),
 $\ell(\sigma_1)=w_1\preceq w$ and $\exists \hat X_{jr}\in \gamma_j$, $\delta({\cal M}_b,w_1)=\hat X_{jr}$.
 By assumption, $\hat X_{jr}\in {\cal X}_m$. Therefore, $\delta({\cal M}_b,w_1)=\hat X_{jr}\in {\cal X}_m$. By Proposition~\ref{prop:Xm}, thus ${\cal C}(w_1)\cap {\cal M}_c \neq \emptyset \Rightarrow |{\cal C}(w_1)|=1$.
\eprof


In other words, an unambiguous cycle $\gamma_j\in \Gamma$, whose observation is $e_{j1}\ldots e_{jk}$, in $B_o$ corresponds to a set of infinite transition sequence $\sigma$ such that $\ell(\sigma)=w=w_0(e_{j1}\ldots e_{jk})^*w_2$ with $w_0\in E^*$ and $w_2\in \{e_{j1},\ldots,e_{jk}\}^*$, and there exists a finite number $K$ and for all $w_1\preceq w$ with $|w_1|>K$, ${\cal C}(w_1)\cap {\cal M}_c \neq \emptyset \Rightarrow |{\cal C}(w_1)|=1$ holds. Namely, the current marking after $w_1$ observed and the subsequent marking after $w_1w'_1\preceq w$ with $w'_1\in E^*$ observed can be determined uniquely if ${\cal C}(w_1w'_1)$ contains a crucial marking.

\prop\label{prop:p-unamb}
Let $G=(N,M_0,\allowbreak E,\ell)$ be an LPN system, and ${\cal M}_c$ the set of crucial markings. There exists a semi-unambiguous cycle $\gamma_j$ in the observer of its BRG iff there exists a finite $K\in \mathbb{N}$ such that $\exists \sigma\in L^\omega(G)$ with $\ell(\sigma)=w$ and $w$ containing the observation of $\gamma_j$, $\forall \sigma'\preceq \sigma$, the condition $\exists \sigma''\in T^*, \ell(\sigma'\sigma'')=w': \sigma'\sigma''\preceq \sigma, |\ell(\sigma'')|<K, {\cal C}(w')\cap {\cal M}_c \neq \emptyset  \Rightarrow |{\cal C}(w')|=1$ holds.
\eprop
\prof
(If) Assume $\exists K\in \mathbb{N}$, $\exists \sigma\in L^\omega(G)$ such that $\forall \sigma'\preceq \sigma$, $\exists \sigma''\in T^*, \ell(\sigma'\sigma'')=w': \sigma'\sigma''\preceq \sigma, |\ell(\sigma'')|<K, {\cal C}(w')\cap {\cal M}_c \neq \emptyset  \Rightarrow |{\cal C}(w')|=1$.
Since $\sigma$ is of an infinite length and $B_o$ has a finite number of nodes, the path along $\ell(\sigma)=w$ must contain a cycle $\gamma_j=\hat X_{j1}e_{j1}\hat X_{j2}\ldots \hat X_{jk}\allowbreak e_{jk}\hat X_{j1}\in \Gamma$, i.e., there exist $w_0\in E^*$ such that $w=w_0(e_{j1}\ldots e_{jk})^*$ with $|w_0|$ is finite. Let $|w_0(e_{j1}\ldots e_{jk})| = K-1 \in\mathbb{N}$. Since $|\ell(\sigma'')|<K$, thus there must exist a state $\hat X_{jr}\in \gamma_j$, $\hat X_{jr}=\delta({\cal M}_b,w')$.
Since ${\cal C}(w')\cap {\cal M}_c \neq \emptyset \Rightarrow |{\cal C}(w')|=1$, by Proposition~\ref{prop:Xm}, thus $\delta({\cal M}_b,w')\in {\cal X}_m$. Therefore, $\exists X_{jr}\in \gamma_j$, $\hat X_{jr} =\delta({\cal M}_b,w_1) \in{\cal X}_m$. Thus the cycle $\gamma_j$ is a semi-unambiguous cycle.

(Only if) Assume that exists a semi-unambiguous cycle $\gamma_j=\hat X_{j1}e_{j1}\hat X_{j2}\ldots \hat X_{jk}\allowbreak e_{jk}\hat X_{j1}\in \Gamma: \exists \hat X_{jr}\in \gamma_j$, $\hat X_{jr}\in {\cal X}_m$. Clearly, there exist $\sigma\in L^\omega(G)$ and $w_0\in E^*$ such that $\ell(\sigma)=w=w_0(e_{j1}\ldots e_{jk})^*$ with $|w_0|$ is finite. Let $|w_0(e_{j1}\ldots e_{jk})| = K-1 \in\mathbb{N}$. Since the $T_u$-induced subnet is acyclic, such that for all $\sigma_1\in L(G)$ with $\ell(\sigma_1)=w_1$, there exists $\sigma_2\in T^*$ with $\ell(\sigma_2)=w_2$, $\delta({\cal M}_b,w_1w_2)=\hat X_{jr}$, and $|w_2|\leq |w_0(e_{j1}\ldots e_{jk})|$(i.e.,$|w_2|< K$). In other words, from any state along $w$ in $B_o$ the state $\hat X_{jr}$ can be always reached within $K$ steps. By assumption, $\hat X_{jr}\in {\cal X}_m$, thus $\delta({\cal M}_b,w_1w_2)=\hat X_{jr}\in {\cal X}_m$. Let $w'=w_1w_2$, that $\delta({\cal M}_b,w')=\hat X_{jr}\in {\cal X}_m$, by Proposition~\ref{prop:Xm}, therefore ${\cal C}(w')\cap {\cal M}_c \neq \emptyset \Rightarrow |{\cal C}(w')|=1$.
\eprof

In simple words, a semi-unambiguous cycle $\gamma_j$ in $B_o$ corresponds to a set of infinite transition sequences $\sigma$ whose observation contains $(e_{j1}\ldots e_{jk})^*$ and for any prefix of $\sigma$ after maximally $K$ observable events the current crucial marking can be uniquely determined. Since the observer is finite, the current crucial marking is detectable periodically. Propositions~\ref{prop:p-unamb} (resp., Proposition~\ref{prop:unamb}) gives the relation between semi-unambiguous (resp., unambiguous) cycles and the estimation of crucial markings. Based on them, sufficient and necessary conditions for C-detectability are derived.

%
%
%

\them\label{therm:S-C-detect}
Let $G$ be an LPN system, ${\cal M}_c$ a set of crucial markings, and $B_o=({\cal{X}},E,\delta,\hat{X}_0,{\cal X}_m)$ the observer of its BRG. LPN $G$ is strongly C-detectable w.r.t ${\cal M}_c$ iff for any $\hat X\in {\cal X}$ reachable from a cycle in $B_o$, $\hat X\in {\cal X}_m$.
\ethem
\prof
(If) Assume that all states reachable from a cycle in $B_o$ is in ${\cal X}_m$. Let $\sigma\in L^\omega(G)$. Since the system is bounded, the observation $w$ of $\sigma$ contains at least a cycle $w_1^* \in E^*$ that corresponds to the observation of a cycle $\gamma_j$ in $B_o$. Therefore, the observation of $\sigma$ can be written as $w=w_0w_1^*w_2$ and $|w_0|$ is finite, where $w_0,w_2\in E^*$. Let $|w_0|=K$, for any prefix $\sigma'$ of $\sigma$, with $\ell(\sigma')=w_0w'$, $|\ell(\sigma')|\geq K$, where $w'\preceq w_1^*w_2$. Let $\hat X=\delta (\hat X_0,w_0w')$. Namely, $\hat X$ is reachable from $\gamma_j$. By assumption that all the states reachable from a cycle in $B_o$ is marked, $\hat X\in {\cal X}_m$. By Proposition~\ref{prop:Xm}, therefore ${\cal C}(w_0w')\cap {\cal M}_c \neq \emptyset \Rightarrow |{\cal C}(w_0w')|=1$.
Therefore, the crucial current marking of $w_0w'$ can also be certainly determined. Analogously, for all the infinite transition sequences their crucial current markings can be determined. Since the system is bounded, there exists a upper bound of $K$ for $w_0w'$ entering a cycle or reaching from a cycle of $B_o$. Therefore, the system is strongly C-detectable.

%
(Only if)
Assume there exist $\gamma_j\in \Gamma$, $\hat X_{jr}\in \gamma_j$, $w'\in E^*$ such that $\delta(\hat X_{jr},w')$ is defined but $\delta(\hat X_{jr},w')\notin {\cal X}_m$.
Clearly, there exists $\sigma\in L^\omega(G)$ and $w=\ell(\sigma)$, $w_1,w_2\in E^*$ such that $w=w_1(e_{j1}e_{j2}\ldots e_{jk})^*w_2$ and $|w_1|$ is finite.
Since the $T_u$-induced subnet is acyclic, such that there exist $\sigma'\preceq\sigma$ with $\ell(\sigma')=w_1(e_{j1}e_{j2}\ldots e_{jk})^*(e_{j1}e_{j2}\ldots e_{jr})w'$, $|\ell(\sigma')|\geq K$($\forall K\in \mathbb{N}$), where $w'\preceq (e_{jr+1} \ldots e_{jk})(e_{j1}e_{j2}\ldots e_{jk})^*w_2$, and states in $\delta({\cal M}_b,w_1(e_{j1}e_{j2}\ldots e_{jk})^*(e_{j1}e_{j2}\ldots e_{jr}))=\hat X_{jr}$. Let $w_0=w_1(e_{j1}e_{j2}\ldots e_{jk})^*(e_{j1}e_{j2}\ldots e_{jr})$, thus $\delta({\cal M}_b,w_0)=\hat X_{jr}$
By assumption that $\exists w'\in E^*$ that $\delta(\hat X_{jr},w')$ is defined, $\delta(\hat X_{jr},w')\notin {\cal X}_m$, thus $\delta({\cal M}_b,w_0w')=\delta(\hat X_{jr},w')\notin {\cal X}_m$.
By Proposition~\ref{prop:Xm}, therefore ${\cal C}(w_0w')\cap {\cal M}_c \neq \emptyset \Rightarrow |{\cal C}(w_0w')|\neq1$.
\eprof

In words, an LPN is strongly C-detectable iff any state reachable from a cycle in the observer is a marked state. It is known that the complexity of finding all the cycles in a directed graph is NP-complete. However, finding all the trongly connected components (SCC) is of polynomial complexity w.r.t the size of the graph. Clearly, if a state of the observer is reachable from a cycle, it is also reachable from an SCC. Therefore, Theorem~\ref{therm:S-C-detect} can be rephrased as the following corollary.

\coro
Let $G$ be an LPN system, ${\cal M}_c$ a set of crucial markings, and $B_o=({\cal{X}},E,\delta,\hat{X}_0,{\cal X}_m)$ the observer of its BRG. LPN $G$ is strongly C-detectable w.r.t ${\cal M}_c$ iff for any $\hat X\in {\cal X}$ reachable from an SCC in $B_o$, $\hat X\in {\cal X}_m$.
\ecoro

\them\label{therm:W-C-detect}
Let $G$ be an LPN system, ${\cal M}_c$ a set of crucial markings, and $B_o$ the observer of its BRG. LPN $G$ is weakly C-detectable w.r.t ${\cal M}_c$ iff in $B_o$ there exists a cycle $\gamma_j$ that is unambiguous w.r.t ${\cal M}_c$.
\ethem
\prof
Follows from Proposition~\ref{prop:unamb}.
%
\eprof

Note that different from that for strong C-detectability, the sufficient and necessary condition for weak C-detectability cannot be reduced to the property of the SCC. The existence of an unambiguous cycle does not imply the existence of such an SCC. However, the converse is true. Thus, we have the following sufficient condition for weak C-detectability.

\coro
Let $G$ be an LPN system, ${\cal M}_c$ a set of crucial markings, and $B_o=({\cal{X}},E,\delta,\hat{X}_0,{\cal X}_m)$ the observer of its BRG. LPN $G$ is
weakly C-detectable w.r.t ${\cal M}_c$ if there exists an unambiguous SCC in $B_o$.
\ecoro

\them\label{therm:PS-C-detect}
Let $G$ be an LPN system, ${\cal M}_c$ a set of crucial markings, and $B_o$ the observer of its BRG. LPN $G$ is periodically strongly C-detectable w.r.t ${\cal M}_c$ iff for all cycles $\gamma_j$ in $B_o$, $\gamma_j$ is semi-unambiguous w.r.t ${\cal M}_c$.
\ethem
\prof
Follows from Proposition~\ref{prop:p-unamb}.
\eprof

We point out that all the SCC are semi-unambiguous does not imply that every cycle is but at leas one. Therefore, the condition for periodically strong C-detectability cannot reduce to the property of the SCC.

\them\label{therm:PW-C-detect}
Let $G$ be an LPN system, ${\cal M}_c$ a set of crucial markings, and $B_o$ the observer of its BRG. LPN $G$ is periodically weakly C-detectable w.r.t ${\cal M}_c$ iff there exits a cycle $\gamma_j$ in $B_o$ such that $\gamma_j$ is semi-unambiguous w.r.t ${\cal M}_c$.
\ethem
\prof
Follows from Proposition~\ref{prop:p-unamb}.
%
\eprof

If there is one semi-unambiguous cycle there is also a semi-unambiguous SCC, and the converse is also true. Thus, we have the sufficient and necessary condition in Corollary~

\coro\label{cor:PWC-D}
Let $G$ be an LPN system, ${\cal M}_c$ a set of crucial markings, and $B_o=({\cal{X}},E,\delta,\hat{X}_0,{\cal X}_m)$ the observer of its BRG. LPN $G$ is
periodically weakly C-detectable w.r.t ${\cal M}_c$ iff there exists an semi-unambiguous SCC in $B_o$.
\ecoro

\coro
Let $G$ be an LPN system, ${\cal M}_c$ a set of crucial markings, and $B_o$ the observer of its BRG. If all cycles in $B_o$ are ambiguous, $G$ does not satisfy any C-detectability w.r.t ${\cal M}_c$.
\ecoro

Theorems~\ref{therm:S-C-detect} to \ref{therm:PW-C-detect} show that, rather than enumerating all reachable markings and constructing the observer of the RG, all four types of C-detectability can be verified through the observer of the BRG. In \cite{shu2011generalized}, a structure called \emph{detector} is proposed to verify strong detectability and periodically strong detectability. Even though the size of the detector is polynomial w.r.t the number of states of the system, it cannot be used to verify weak and periodically weak detectability. The observer of the BRG can be used to to verify all the C-detectability and when the set of crucial markings is changed, one only needs to update the value of functions $\alpha$ and $\beta$.

In the next section, we show that when the set of crucial markings is described by a set of linear constraints, the value of $\alpha$ and $\beta$ functions can be determined through solving the integer linear programming problems.
\subsection{${\cal M}_c$ described by GMECs}

%
It is well-known that the general mutual exclusive constraints (GMECs) \cite{giua92} describe interesting subsets of the state space of a net and provide a linear algebra tool for Petri net analysis. To simplify the problem, now we assume that the set of crucial markings is described by a set of GMECs $${\cal M}_c=(W,K)=\bigcap_{i=1}^r\{M\in \mathbb{N}^m|w_i^T\cdot M\leq k_i\},$$ where $w_i \in \mathbb{Z}^m$ and $k_i\in \mathbb{Z}$ with $i=1,2,\cdots,r$. Such a set of GMECs $(w_i,k_i)$ is denoted as ${\cal M}_c=\{M\in \mathbb{N}^m|W\cdot M\leq K\}$, where $W=[w_1,w_2,\cdots,w_r]^T$ and $K=[k_1,k_2,\cdots,k_r]^T$. In addition, the following constraint set is also defined.

\dfn\label{def:constraint}
Let $M\in R(N,M_0)$ be a marking of an LPN system $G=(N,M_0,E,\ell)$, ${\cal M}_c=\{M\in \mathbb{N}^m|W\cdot M\leq K\}$ the set of crucial markings.
\begin{itemize}
  \item The \emph{${\cal Y}(M)$-constraint set} is defined by
\begin{equation}\label{eq:Y}
{\cal Y}(M)=\left\{ \begin{aligned}
                       & M'=M+C_u\cdot y\\
                       & W^T\cdot M \leq K\\
                       & M' \in \mathbb{N}^{m}\\
                       & y \in \mathbb{N}^{n_u}\\
                          \end{aligned} \right.
\end{equation}
  \item The \emph{${\cal Z}(M)$-constraint set} is defined by
\begin{equation}\label{eq:Z}
{\cal Z}(M)=\left\{ \begin{aligned}
                       &M'=M+C_u\cdot y\\
                       &M' \in \mathbb{N}^{m}\\
                       &y \in \mathbb{N}^{n_u}_{\geq 1}\\
                          \end{aligned} \right.
\end{equation}
\end{itemize}
\hfill $\diamond$
\edfn

By Eqs.~\eqref{eq:alpha}, \eqref{eq:beta} and \eqref{eq:ur}, the following results hold.
\prop\label{prop:mCurrent}
Given a basis marking $M_b\in {\cal M}_b$ of an LPN system $G$,
\begin{itemize}
  \item $\alpha(M_b)=1$ iff the ${\cal Y}(M_b)$-constrain set is feasible;
  \item $\beta(M_b)=1$ iff the ${\cal Z}(M_b)$-constrain set is feasible.
\end{itemize}
\eprop


Based on Proposition~\ref{prop:mCurrent}, the construction of the BRG for C-detectability requires solving integer linear programming problems (ILPP). However, for some net structures (see \cite{tong2017verification}) the complexity of constructing the BRG can be reduced by relaxing ILPP into linear programming problems (LPP).

\exm
Consider again the LPN system in Fig.~\ref{fig:LPN}. Let the set of crucial markings be ${\cal M}_c=\{M\in \mathbb{N}^4|M(p_1)\geq 1\}$, i.e., $W=[\begin{array}{cccc}
  -1 & 0 & 0 & 0\\
\end{array}]^T$ and $K=-1$. By solving Eq.~\eqref{eq:Y}, $\alpha(M_0)=1$ and $\alpha(M_1)=0$. By solving Eq.~\eqref{eq:Z}, $\beta(M_0)=0$ and $\beta(M_1)=1$.
Therefore, the obtained BRG for C-detectability is identical to the one in Fig.~\ref{fig:BRG} and the observer is identical to the one in Fig.~\ref{fig:observer} and $\mathcal{X}_m=\{\hat{X}_2\}$.

There is no unambiguous cycle in the observer. By Theorems~\ref{therm:S-C-detect} and \ref{therm:W-C-detect}, the LPN system is neither strongly detectable nor weakly detectable w.r.t ${\cal M}_c$. On the other hand, there is a cycle $\gamma_1=\hat{X}_2 a \hat{X}_1 b \hat{X}_2$ containing the marked states $\hat{X}_2$. By Definition~\ref{def:unamb}, $\gamma_1$ is semi-unambiguous. Therefore, by Theorem~\ref{therm:PW-C-detect} the LPN is periodically weakly C-detectable w.r.t ${\cal M}_c$. However, there also exists a cycle $\gamma_2= \hat{X}_2 a \hat{X}_2$ that is not semi-unambiguous, and by Theorem~\ref{therm:PS-C-detect} the LPN system is not periodically strongly detectable. \hfill $\diamond$
\eexm

%

%
\section{Conclusions and future work}
In this paper, a novel approach to verifying C-detectability of labeled Petri nets is developed. We show that the notions of basis markings can also be effectively applied to verifying C-detectability. A BRG for C-detectability is proposed. For Petri nets whose unobservable subnet is acyclic, the C-detectability property can be decided by just constructing the observer of the BRG that is usually much smaller than the RG. The future research is to use the verifier net to further reduce the complexity of verifying strong C-detectability and Periodically strong C-detectability properties.

\section*{Acknowledgment}
This work was supported by the National Natural Science Foundation of China under Grant No. 61803317,
the Fundamental Research Funds for the Central Universities under Grant No. 2682018CX24,
and the Sichuan Provincial Science and Technology Innovation Project under Grant No. 2018027.

\ifCLASSOPTIONcaptionsoff
  \newpage
\fi



%
%
%

\bibliographystyle{IEEEtran}
\bibliography{C-detectability}

%
%
%
%
%
%
%
%
%
\end{document}